\documentclass[a4paper,11pt]{article}
\pdfoutput=1

\usepackage{jcappub} 

\usepackage[T1]{fontenc} 

\title{Matter power spectrum and the challenge of percent accuracy}

\author[a]{Aurel Schneider,}
\author[a]{Romain Teyssier,}
\author[a]{Doug Potter,}
\author[a]{Joachim Stadel,}
\author[b]{Julian Onions,}
\author[a]{Darren S. Reed,}
\author[c]{Robert E. Smith,}
\author[d,e]{Volker Springel,}
\author[b]{Frazer R. Pearce,}
\author[f]{and Roman Scoccimarro}

\affiliation[a]{Institute for Computational Science, University of Zurich, Winterthurerstrasse 190, 8057 Zurich, Switzerland}
\affiliation[b]{School of Physics and Astronomy, University of Nottingham, Nottingham NG7 2RD, United Kingdom}
\affiliation[c]{Department of Physics and Astronomy, University of Sussex, Brighton, BN1 9QH, United Kingdom}
\affiliation[d]{Heidelberger Institut f\"ur Theoretische Studien, 69118 Heidelberg, Germany}
\affiliation[e]{Zentrum f\"ur Astronomie der Universit\"at Heidelberg, ARI, 69120 Heidelberg, Germany}
\affiliation[f]{Center for Cosmology and Particle Physics, Department of Physics, New York University, NY 10003, New York, USA}

\emailAdd{aurel@physik.uzh.ch}

\abstract{Future galaxy surveys require one percent precision in the theoretical knowledge of the power spectrum over a large range including very nonlinear scales. While this level of accuracy is easily obtained in the linear regime with perturbation theory, it represents a serious challenge for small scales where numerical simulations are required. In this paper we quantify the precision of present-day $N$-body methods, identifying main potential error sources from the set-up of initial conditions to the measurement of the final power spectrum. We directly compare three widely used $N$-body codes, {\tt Ramses}, {\tt Pkdgrav3}, and {\tt Gadget3} which represent three main discretisation techniques: the particle-mesh method, the tree method, and a hybrid combination of the two. For standard run parameters, the codes agree to within one percent at $k\leq1$ $h\,\rm Mpc^{-1}$ and to within three percent at $k\leq10$ $h\,\rm Mpc^{-1}$. We also consider the bispectrum and show that the reduced bispectra agree at the sub-percent level for $k\leq 2$ $h\,\rm Mpc^{-1}$.
In a second step, we quantify potential errors due to initial conditions, box size, and resolution using an extended suite of simulations performed with our fastest code {\tt Pkdgrav3}. We demonstrate that the simulation box size should not be smaller than $L=0.5$ $h^{-1}\rm Gpc$ to avoid systematic finite-volume effects (while much larger boxes are required to beat down the statistical sample variance). Furthermore, a maximum particle mass of $M_{\rm p}=10^{9}$ $h^{-1}\rm M_{\odot}$ is required to conservatively obtain one percent precision of the matter power spectrum. As a consequence, numerical simulations covering large survey volumes of upcoming missions such as {\tt DES}, {\tt LSST}, and {\tt Euclid} will need more than a trillion particles to reproduce clustering properties at the targeted accuracy.
}

\begin{document}
\maketitle
\flushbottom


\section{Introduction}
In the last decades, cosmology has entered the high precision regime owing to ever more accurate measurements of the Cosmic Microwave Background (CMB). The statistical information of the CMB sky is, however, intrinsically limited, while large scale structures contain a great wealth of modes which can be exploited, provided nonlinear structure formation is well understood. Next-generation galaxy and weak lensing surveys such as {\tt DES}\footnote{\tt www.darkenergysurvey.org}, {\tt LSST}\footnote{\tt www.lsst.org/lsst}, and {\tt Euclid}\footnote{\tt sci.esa.int/euclid} require percent accurate modelling of the matter power spectrum up to wave numbers of $k\sim10$ $h\,\rm Mpc^{-1}$ in order to fully exploit their constraining power for cosmology \citep{Ivezic2008,Laureijs2011}.

Standard perturbation theory gives accurate results up to $k\sim0.1$ $h\,\rm Mpc^{-1}$, while numerical simulations are indispensable at higher wave numbers \citep{Carlson2009,Fosalba2015}. Pure dark matter simulations based on $N$-body techniques are believed to be accurate up to about $k\sim0.5$ $h\,\rm Mpc^{-1}$, beyond which baryonic feedback from active galactic nuclei (AGN) must be included \citep{vanDaalen2011, Schneider2015}.

In this paper we focus on the matter power spectrum from collisionless $N$-body simulations, ignoring all hydrodynamical effects. Although strictly not valid at small scales, this approach is currently the only option for precision cosmology as the relevant AGN feedback mechanism is not well understood, and is poorly constrained by observations. A potential way forward is to study AGN feedback with high resolution simulations of small cosmological volumes and to parametrise the effects on the matter power spectrum. Cosmological parameter estimation can then be carried out on the basis of $N$-body simulations plus additional free model parameters accounting for the AGN contribution \citep{Mohammed2014, Schneider2015}.

Comparison studies of $N$-body codes and subsequent analysis tools have been performed in the past. The first investigations of different $N$-body techniques was carried out in Ref.~\citep{Efstathiou1985} more than thirty years ago. More recently, the authors of Ref.~\citep{OShea2005} compared high-redshift power spectra and halo abundances from mesh- and particle-based techniques, reporting significative differences at small scales. The first detailed code comparison study including six gravity codes was carried out by Ref.~\citep{Heitmann2005}. In terms of the power spectrum, the authors reported agreement of roughly ten percent between particle codes up to $k\sim 10$, while mesh codes deviated already at smaller wave-numbers due to incomplete resolution. Three years later, another comparison project with 10 gravity codes was carried out by Ref.~\citep{Heitmann2008}, showing further improvement in code agreement and stating roughly one percent differences for the power spectrum below $k\sim 1$ h/Mpc. At larger wave-numbers they observed growing discrepancy between mesh and particle codes which exceeded ten percent at $k\sim 10$ h/Mpc, while both methods were shown to agree to about five percent among each other. Finally, Ref.~\citep{Heitmann2010} reinvestigated the difference between mesh and particle methods using simulations with larger, cosmological viable box sizes and particle numbers. They confirmed the reported percent agreement up to $k=2$ h/Mpc.

In the present paper we build upon these past efforts and compare three gravity codes representing the most widely used discretisation techniques. We thereby use an unprecedented setup in terms of box size and particle resolution to allow for a code comparison free of systematic effects. This is confirmed in the second part of the paper were we show that potential systematics from initial conditions, box size, and particle numbers are below the percent error condition.

Sec.~\ref{codecomp} is devoted to the code comparison, focusing on the auto and cross power spectra. In Sec.~\ref{Nbodytest} we take a critical look at the simulation pipeline and investigate the accuracy of the initial conditions as well as potential finite volume and resolution effects. A summary of the results including a list of requirements to obtain percent accuracy of the matter power spectrum is presented in Sec.~\ref{Conclusions}. In the Appendices, we investigate modifications of the code parameters (Appendix \ref{ApdxA}) and we present a code comparison beyond the power spectrum (Appendix \ref{ApdxB}).


\section{Code comparison}\label{codecomp}
The first part of this paper is about comparing $N$-body codes with respect to the precision requirements of upcoming galaxy and weak lensing surveys. Our study is mainly focused on the auto power spectrum which is the prime statistical measure in cosmology. Additionally, we investigate phase-shifts in Fourier space by cross-correlating the results of different codes in order to further quantify the spatial disagreement between density fields from different $N$-body techniques. 

\subsection{$N$-body codes}
The gravitational $N$-body technique is the standard tool to simulate the nonlinear Universe, yielding accurate results at scales where hydrodynamical effects are subdominant. Most $N$-body codes are either based on a particle-mesh method, a tree algorithm, or a hybrid combination of the two. In this paper, we compare the codes {\tt Ramses}, {\tt Pkdgrav3}, and {\tt Gadget3}, which represent each of these three approaches and are widely used in the astrophysics and cosmology community.

The comparison is performed by running a simulation of box size $L=500$ $h^{-1}\rm Mpc$ and resolution of $N=2048$ particles per dimension with each of the three codes, starting from the exact same initial conditions and using the standard run parameters described below. The initial conditions are based on second order Lagrangian perturbation theory (2LPT), generated at redshift 49 with a modified version of the IC code from \citep{Crocce2006,Scoccimarro2012}. For the cosmological parameters, we use Planck values, i.e., $\Omega_m=0.3071$, $\Omega_{\Lambda}=0.6929$, $\Omega_b=0.0483$, $h=0.6777$, $n_s=0.9611$, and $\sigma_8= 0.8288$ \citep{Ade2014}. The measurement of the power spectra is performed at exactly the same redshifts and with the same analysis tool (using the triangular shaped cloud scheme for the mass assignment). In this way, we carefully avoid all other potential sources of error and directly compare effects due to the gravity calculations of the codes.

We now briefly present the three codes and give details about the run parameters for the comparison:
\begin{itemize}
\item The $N$-body and hydrodynamical code {\tt Ramses} \citep{Teyssier2002} is based on a particle-mesh technique and uses adaptive mesh refinement for high density regions. The code is mainly used for hydrodynamical simulations in a cosmological context \citep{Ocvirk2008, Agertz2011} but it has also been employed for pure dark matter $N$-body runs in the past \citep{Teyssier2009,Rasera2014}. For the comparison, we apply a coarse-level grid with refinement level $\ell_{\rm min}=12$, corresponding to $2048^3$ coarse cells. New refinements are triggered on a cell-by-cell, recursive basis when a cell collects more than 8 particles. Using this strategy we reach a maximum level of refinement $\ell_{\rm max}=18$, corresponding to a spatial resolution of 2 $h^{-1}\rm kpc$. We employ adaptive, level-by-level time-stepping, with a time step size set smaller than the local free fall time, and by the requirement that a particle cannot move more than half a cell within one time step. The convergence criterion for the Poisson solver, defined as the ratio of the residual $L^2$-norm to the right-hand side $L^2$-norm, is set to $\epsilon=10^{-4}$.
\item The gravity code {\tt Pkdgrav3} \citep[an earlier version of which is described in][]{Stadel2001} is based on a binary tree algorithm using fifth order fast multipole expansion of the gravitational potential (using cell-cell interactions making it an $\mathcal O(N)$ gravity calculation method). Periodic boundaries conditions are calculated with the Ewald summation technique, requiring very little data movement while exposing a high degree of parallelism. {\tt Pkdgrav} has been extensively used for $N$-body simulations in the past, mainly in the context of cosmological zoom simulations such as {\it Via Lactea} \citep{Diemand2007} and {\it Ghalo} \citep{Stadel2009}. The current version of {\tt Pkdgrav} includes GPU acceleration for all force calculations, leading to a significant speed-up with respect to previous versions. In this paper, we use the run parameters $\varepsilon=0.02 \,l_{\rm mean}$ (where $l_{\rm mean}$ is the mean particle separation) and $\theta=0.7$ ($\theta=0.55$ above redshift two) for softening and tree opening criteria. The adaptive time-stepping is parametrised in the standard way, i.e $dt_i = \eta\sqrt{\varepsilon/|a_i|}$ with $\eta=0.15$ (where $a_i$ is the acceleration of particle $i$). {\tt Pkdgrav3} also has a more sophisticated time-stepping criterion based on an estimation of the local dynamical time. 
\item The tree-particle-mesh code {\tt Gadget3} applies a uniform particle-mesh method at large scales plus a first order oct-tree technique at small scales \citep[see][for a description of an earlier version of the code]{Springel2005}. {\tt Gadget} is extensively used in many contexts and is most known for the {\it Millennium} suite of cosmological simulations \citep{Springel2005b}, as well as the {\it Aquarius} zoom simulations \citep{Springel2008}. For the comparison, we use a comoving Plummer-equivalent softening length of $\varepsilon=10$  $h^{-1}\rm kpc$ and the code's relative tree opening criterion with a tolerance value of $\alpha = 0.005$ for the gravitational force accuracy \citep[see][for more information]{Springel2005}. Furthermore, we adopt a time integration parameter corresponding to $\eta = 0.22$ for the time-stepping criterion used above in {\tt Pkdgrav3}. The long-range particle-mesh forces are calculated with a $2048^3$ Fourier grid.
\end{itemize}

All simulations used of this comparison were performed on the hybrid CPU/GPU cluster {\it Piz Daint} at the Swiss National Supercomputing Centre (CSCS). The total run-time for the three codes is $94\,352$ node-hours for {\tt Ramses}, $34\,524$ node-hours for {\tt Gadget3}, and $1632$ node-hours for {\tt Pkdgrav3} (the former two codes were run on 512 nodes using CPU only, while the latter was run on 128 nodes using full GPU acceleration for the force calculations). On each node of {\it Piz Daint} there are 8 CPU-cores and one Nvidia Tesla K20X GPU accelerator.

\subsection{Definitions}
Before discussing differences between the gravity codes, we give a brief definition of the power spectrum and cross power coefficient (definitions of the propagator and the bispectrum can be found in Appendix \ref{ApdxB}). Let us first assume we have two density fields $\delta_X(\mathbf{k})$ and $\delta_Y(\mathbf{k})$ in Fourier space. The power spectrum $P_{XY}(k)$ is then defined as (see e.g. Ref.~\citep{Bernardeau2002})
\begin{equation}\label{powspec}
\langle\delta_X(\mathbf{k})\delta_Y(\mathbf{k}') \rangle\equiv \delta_D(\mathbf{k}+\mathbf{k}')P_{XY}(k),
\end{equation}
where $\delta_D(\mathbf{x})$ is the three dimensional Dirac delta function. Eq. \eqref{powspec} defines both the auto and the cross power spectrum, which we now briefly discuss. The auto power spectrum is given by
\begin{equation}
P(k)\equiv P_{XX}(k)
\end{equation}
and provides a measure of the density \emph{amplitudes} at different $k$-modes. The cross power spectrum $P_{XY}$ (with $X\neq Y$), on the other hand, also measures the \emph{phase} differences between density fields. It is convenient to the define the cross power coefficient (e.g. \citep{Pen1998,Tegmark1999,Seljak2004})
\begin{equation}\label{phasespectrum}
r_{XY}(k)\equiv \frac{P_{XY}(k)}{\sqrt{P_{XX}(k)P_{YY}(k)}},
\end{equation}
which only contains information about phase shifts while all amplitudes are factored out. This becomes evident if we split the perturbation field into an amplitude and a phase component, i.e. $\delta_X(\mathbf{k})=\Delta_X(\mathbf{k})\exp\left[i\phi_X({\mathbf{k}})\right]$ (see e.g. Ref.~\cite{Chiang2000}). The fact that the cross power coefficient measures the spatial shifts between density fields makes it an interesting alternative indicator for a code comparison.

\subsection{Auto power spectrum}
Analysing the accuracy of the auto power spectrum $P(k)$ from numerical simulations is the main goal of this paper. The measurement of $P(k)$ is performed with fully mpi-parallel Fast Fourier Transform (FFT) using the triangular shaped cloud (TSC) method to assign particles on the grid\footnote{In order to avoid smearing effects, we normalise the density contrast in $k$-space with the Fourier transform of the assignment window (see e.g. \citep{Jing2005,Cui2008,Colombi2009} for more information).}.

\begin{figure}
\centering
\includegraphics[trim = 5mm 5mm 60mm 12mm, clip, scale=1.0]{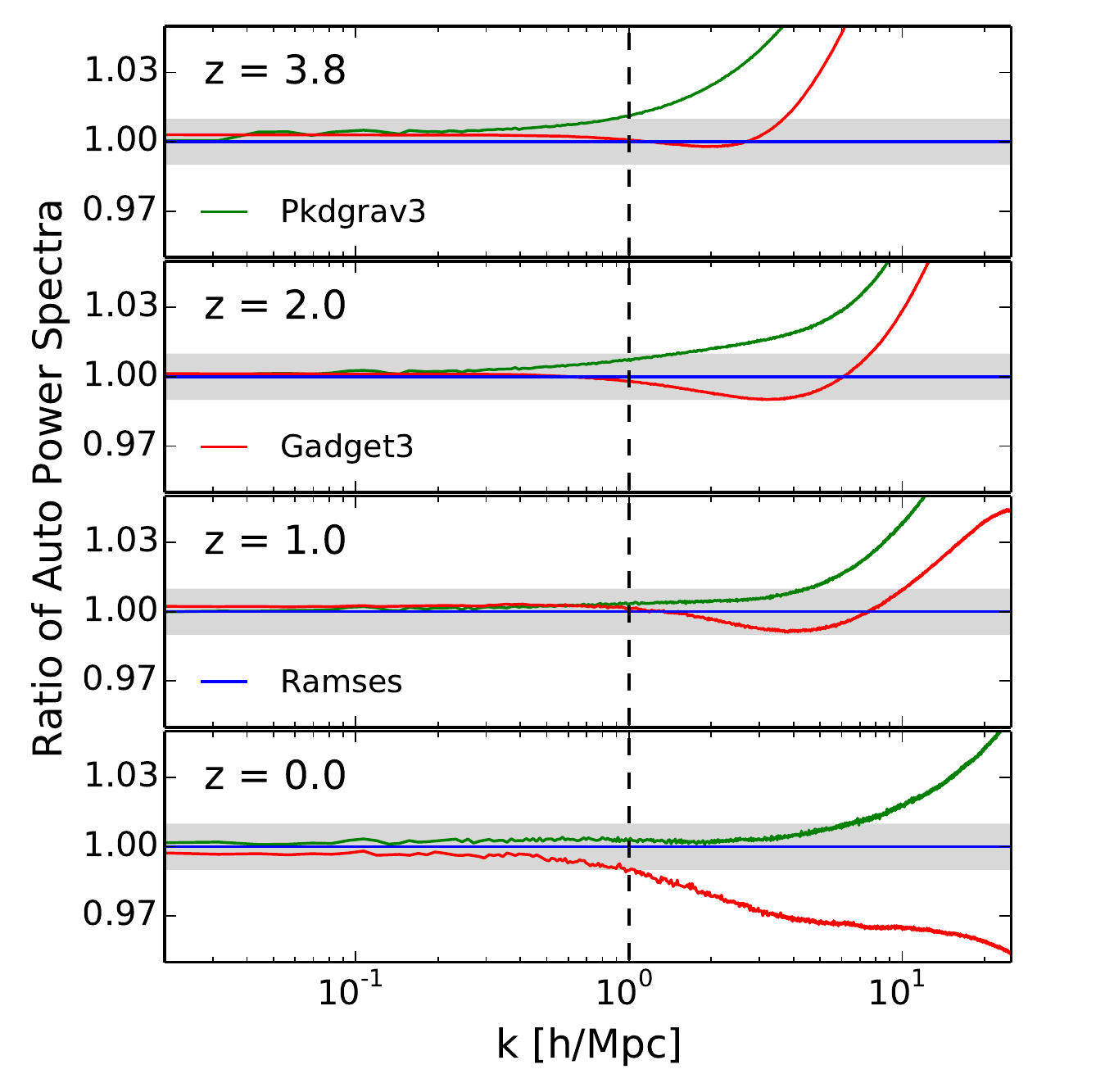}
\caption{\label{fig:PScomp}Comparison of auto power spectra from the three different $N$-body codes at different redshifts. Green lines correspond to {\tt Pkdgrav3}, red lines to {\tt Gadget3}, and blue lines to {\tt Ramses} (reference lines). One percent agreement (indicated by the grey band) is obtained for $k\leq 1$ $h\,\rm Mpc^{-1}$ (dashed vertical line).}
\end{figure}

The resulting power spectra are shown in Fig. \ref{fig:PScomp}, where different panels correspond to different redshifts $z=3.8$, 2, 1.0, and 0.0. The green lines refer to {\tt Pkdgrav3}, the red lines to {\tt Gadget3}, and the blue lines to {\tt Ramses}\footnote{We have chosen the blue lines to act as reference, solely because it lies between the green and red lines at $z=0$, therefore improving the readability of the plots.}. One percent agreement (grey shaded area) between the different codes is obtained up to $k\sim 1$ $h\,\rm Mpc^{-1}$ over all redshifts, as illustrated by the vertical dashed line. In the highly nonlinear regime from $k=1$ to $10$ $h\,\rm Mpc^{-1}$, the agreement between codes is at the three percent level for $z=1$ and below\footnote{While {\tt Ramses} and {\tt Pkdgrav3} show percent agreement until $k\sim7$ $h\,\rm Mpc^{-1}$, {\tt Gadget} slightly deviates at $k\sim1$ $h\,\rm Mpc^{-1}$.}. At higher redshifts, the discrepancy grows, reaching about five percent at $z=2$ and ten percent at $z=3.8$.

The agreement between the codes is significantly better than in previous code comparison projects by \citet[hereafter H08]{Heitmann2008} and \citet[H10]{Heitmann2010} illustrating the progress in code development over the last five years. At very large scales we obtain maximal differences of $\sim0.4$ percent between codes, with respect to $\sim 3$ percent in H08 and $\sim 2$ percent in H10\footnote{The better agreement between {\tt Pkdgrav3} and {\tt Gadget3} is (at least partially) because of a new implementation of the periodic boundary conditions in {\tt Pkdgrav}.}. At small scales beyond $k\sim 1$ h/Mpc, the prominent systematic offset between PM-codes and tree-codes visible in H08 (with more than 10 percent difference at  $k\sim 10$ $h\,\rm Mpc^{-1}$) has now entirely disappeared.

The relatively large difference between {\tt Pkdgrav3}/{\tt Gadget3} and {\tt Ramses} at z=3.8 can be explained by the fact that AMR codes require many more particles to resolve haloes. As we will see in the following sections, the power spectrum is dominated by group-sized haloes at redshift zero, which are well resolved in our simulations. At high redshift, however, the signal stems from considerably smaller structures which are better resolved with the tree-codes than with an AMR technique. Higher resolution of the AMR grid is required to remedy this.

The results of the code comparison are based on the standard run parameters described above. These parameters have never been systematically tested in the context of large-scale cosmological simulations, but they emerged via many different code applications in the past. However, finding more optimal code parameters is non-trivial because the true power spectrum is not known {\it a priori} . Parameters cannot simply be tuned to achieve maximal agreement between codes since this could lead to convergence towards the wrong answer.

It is nevertheless important to quantify the dependency of the run parameters on the resulting power spectrum. Only results which are insensitive to the choice of code parameters can be trusted. In  Appendix \ref{ApdxA} we study the effects of the most sensitive code parameters, which are the size of the PM grid for {\tt Gadget3} as well as the time-stepping criterion for all three codes. We conclude that reasonable variations of these parameters lead to sub-percent effects on the power spectrum below $k\sim10$ $h\,\rm Mpc^{-1}$, smaller than the observed differences between codes visible in Fig.~\ref{fig:PScomp}.

Summing up, the code comparison suggests that the consensus between different $N$-body techniques is good, however not quite good enough for the targeted percent accuracy up to $k\sim10$ $h\,\rm Mpc^{-1}$. Further improvements to the codes will not be easily achievable as the correct solution for the matter power spectrum is not known. A common effort of the community is required to converge towards a generally accepted solution. In order to encourage further comparison of $N$-body codes, we release the IC file used here plus all power spectrum measurements on {\tt www.ics.uzh.ch/{\raise.17ex\hbox{$\scriptstyle\sim$}}aurel/}.

\subsection{Cross power spectrum}
The cross power coefficient $r_{XY}$ (defined in Eq.~\ref{phasespectrum}) quantifies the spatial shifts between two density fields and is therefore a sensitive statistical measure to compare $N$-body codes. While the auto power spectrum only gives information about the amplitude of perturbations, the cross power coefficient measures the relative phase-shifts for any given $k$-mode. The cross power spectrum is obtained by separately Fourier-transforming the two density fields from different $N$-body codes.

\begin{figure}
\centering
\includegraphics[trim = 0mm 40mm 0mm 0mm, clip,scale=0.8]{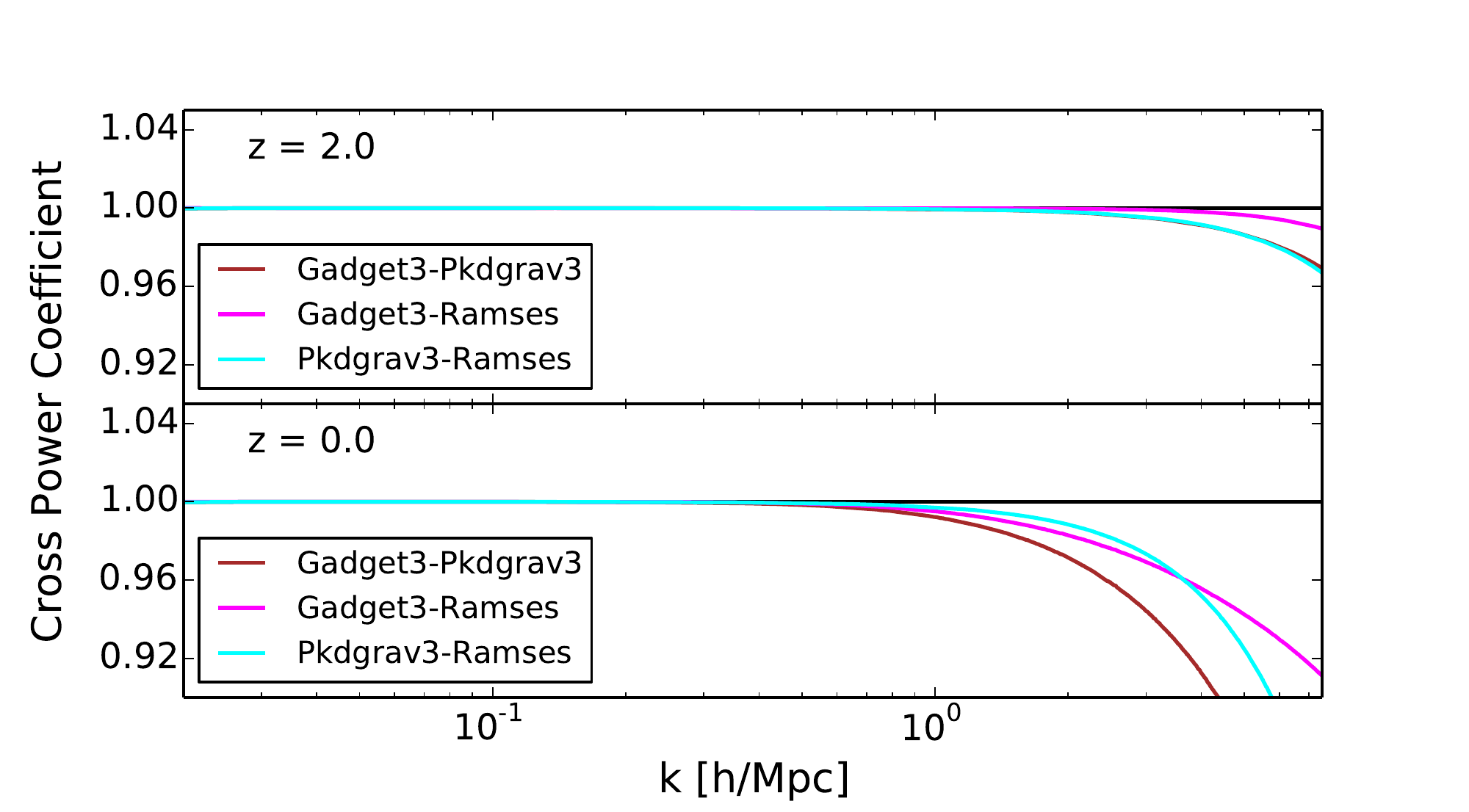}
\caption{\label{fig:phaseS}Cross power coefficient (as defined in Eq.~\ref{phasespectrum}) at redshift two (top) and redshift 0 (bottom). Brown, magenta, and cyan lines correspond to the combinations of density fields from {\tt Gadget3-Pkdgrav3}, {\tt Gadget3-Ramses}, and {\tt Pkdgrav3-Ramses}.}
\end{figure}

In Fig.~\ref{fig:phaseS} we plot the cross power coefficients based on density fields from {\tt Gadget3}-{\tt Pkdgrav3} (brown), {\tt Gadget3}-{\tt Ramses} (magenta), and {\tt Pkdgrav3}-{\tt Ramses} (cyan) at redshift 2 (top) and redshift 0 (bottom). At the largest scales  ($k\lesssim0.5$ $h\,\rm Mpc^{-1}$), the results are in perfect agreement, which can be explained by the fact that the cross power coefficient is independent of the growth factor and therefore insensitive to errors related to global time-integration. At smaller scales and especially at low redshift, the deviations are considerably larger than the differences observed in the power spectrum. This is due to the effect of gravity which magnifies deviations of phases over time, something that is clearly visible in Fig.~\ref{fig:phaseS}.

In general, the cross power coefficients from different code combinations are in good agreement with each other. At redshift 2, there are no visible phase-shifts up to $k\sim2$ h/Mpc. At smaller scales some small differences start to appear, while the density fields from {\tt Gadget3} and {\tt Ramses} seem to agree somewhat better with each other than with the density field from {\tt Pkdgrav3}. At redshift zero, phases-shifts start to be visible above $k\sim 0.5$ $h\,\rm Mpc^{-1}$. The largest differences are observed between the density fields of {\tt Gadget3} and {\tt Pkdgrav3}, which is surprising given the fact that they use similar numerical techniques at small scales.

In summary, we want to highlight the extremely good agreement of the cross power coefficients below $k\sim 0.5$ h/Mpc suggesting vanishing force errors at large scales. As a consequence, the sub-percent differences visible in power spectrum at large scales (see Fig.~\ref{fig:PScomp}) have to come from slightly different growth factors and are therefore potentially stemming from small inaccuracies in the global time-integration schemes. Finally, we want to stress that both the auto and cross power spectra shown in this paper only provide measures for the relative differences between codes but do not indicate which one of the three codes is most accurate.


\section{Testing the $N$-body pipeline}\label{Nbodytest}
Potential inaccuracies of numerical simulations are not restricted to the $N$-body code but can stem from the initial conditions, limited box-size or physical resolution. Each of these sources of error has been extensively studied in the past (see e.g. \citep{Heitmann2010,Rasera2014}). Here, we reanalyse potential effects from initial conditions, box-size and resolution with the focus on the requirement of sub-percent errors. 

\subsection{Initial Conditions}
Initial conditions of cosmological simulations are generated as a random realisation of a (Gaussian) density field, based on either first or second order Lagrangian perturbation theory. The density field is usually discretised in form of aligned particles on an initial grid, where small displacements account for the initial perturbations.

The redshift of the initial conditions has to be chosen with care. It should lie in a range where all resolved perturbations are large enough to dominate numerical noise, but still small enough to be accurately described by perturbation theory. It has been shown in the past that it is advantageous to use second order Lagrangian perturbation theory (2LPT) with respect to the simpler first order or Zel'dovich approximation (ZA), as it allows for smaller starting redshifts, further away form the noise dominated high-redshift regime \citep{Scoccimarro1998,Crocce2006,Jenkins2010,Reed2013}.

\begin{figure}
\centering
\includegraphics[trim = 0mm 40mm 0mm 0mm, clip, scale=0.8]{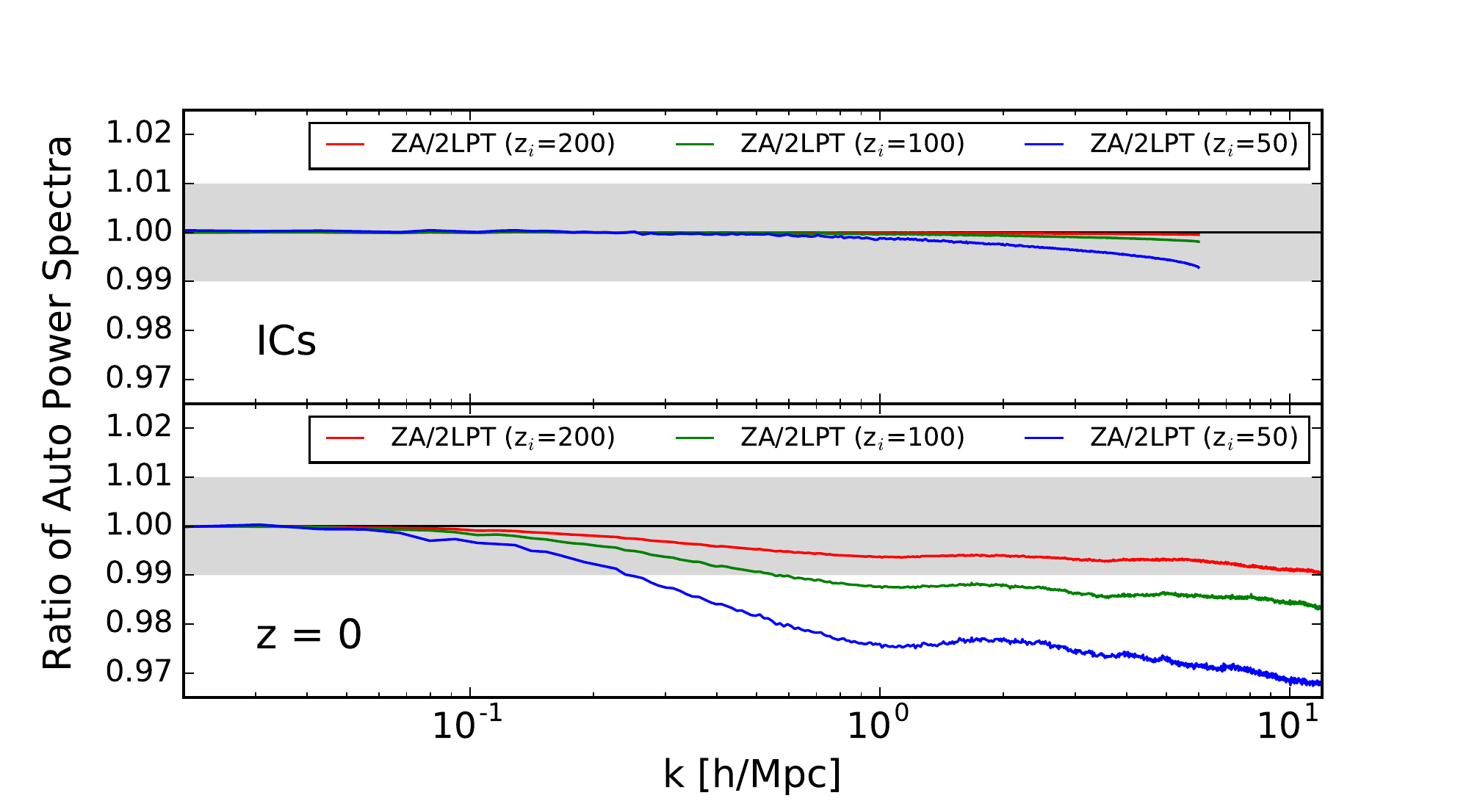}
\caption{\label{fig:PSIC}Ratios of power spectra from simulations with ZA and 2LPT initial conditions and different starting redshifts ($z_i$). The top panel shows measurements at redshift $z_i$, the bottom panel at redshift zero. One percent agreement is illustrated by the grey band.}
\end{figure}

We study the effects of the initial conditions on the power spectrum at redshift zero by running simulations with $L=512$ $h^{-1}\rm Mpc$ and $N=1024$ particles per dimension with the $N$-body code {\tt Pkdgrav3}. The initial conditions are generated with {\tt MUSIC} \citep{Hahn2011}, using both the ZA and 2LPT approach at different starting redshifts ($z_i$). 

The resulting effects on the power spectrum are illustrated in Fig.~\ref{fig:PSIC}. In the top panel, we show the ratios between ZA and 2LPT directly measured in the ICs at the corresponding starting redshifts of $z_i=200$ (red line), $z_i=100$ (green line), and $z_i=49$ (blue line). The differences between ZA and 2LPT are at sub-percent level (converging towards large $z_i$) and limited to high wave numbers above $k\sim1$ $h\,\rm Mpc^{-1}$. In the bottom panel, we show the same ratios now measured at redshift zero. The differences between ZA and 2LPT have grown substantially affecting wave number beyond $k\sim0.1$ $h\,\rm Mpc^{-1}$. This behaviour is in agreement with previous findings \citep{Crocce2006,Knebe2009,L'Huillier2014}.

Fig.~\ref{fig:PSIC} suggests that a starting redshift $z_i\gtrsim200$ is required to obtain percent accuracy with ZA initial conditions. Such high starting redshifts are prone to numerical problems, since $N$-body codes do not deal well with extremely small initial density perturbations. At what redshift numerical effects become a problem depends on the code and the run parameters. Based on a study involving {\tt Pkdgrav2} and {\tt Gadget2}, \citet{Reed2013} concluded that the initial redshift and the redshift of typical halo formation should not differ by more than a factor of fifty. For the cosmological boxes investigated here, most haloes form around redshift two \citep[see for example][]{McBride2009} which results in the requirement $z_i\lesssim 100$\footnote{Recent tests with {\tt Pkdgrav3} show that high-redshift errors can be reduced by choosing a smaller tree-opening parameter during the first gravity steps. This could potentially allow to shift the starting redshift to higher values for the same precision requirements.}.

In agreement with previous results, we conclude that initial conditions with 2LPT should be used consistently for cosmological simulations. They are significantly more accurate than ZA initial conditions and they allow lower starting redshifts, thus decreasing the run-time of simulations.


\subsection{Box size and resolution}
A careful setup of simulations in terms of box size and particle numbers is crucial in order to obtain one percent agreement in the power spectrum. Small boxes tend to suffer from sample variance and missing large-scale modes, while large boxes might not have enough resolution to capture the very nonlinear scales.  

It is straight-forward to determine the expected statistical (Gaussian) error which consists of a sample variance and a shot-noise contribution and is given by
\begin{equation}\label{staterr}
\Delta P(k)=\left(\frac{2}{\Delta N_{\rm m}}\right)^{1/2}\left[P(k)+P_{\rm sn}\right],
\end{equation}
where $\Delta N_{\rm m}=L^3k^2\Delta k/(2\pi^2)$ is the number of modes per $k$-bin and $P_{\rm sn}\equiv(L/N)^3$ is the Poisson shot-noise. From Eq.~\eqref{staterr} it becomes obvious that the sample variance increases towards larger scales. Enforcing sub-percent statistical errors and assuming $\Delta k=2\pi/L$ results in the condition
\begin{equation}
L\gtrsim \frac{250}{k}
\end{equation}
for the box length $L$. This means that a minimal box length of $L=2.5$ $h^{-1}\rm Gpc$ is required to beat down the sample variance below the percent level for $k$-modes above $k\sim 0.1$ $h\,\rm Mpc^{-1}$. The shot noise contribution $P_{\rm sn}$, on the other hand, becomes important at the smallest scales. In this paper we subtract Poisson shot-noise from all measured power spectra and we indicate the scale above which shot-noise contributes at more than one percent to the total power spectrum.

Next to statistical errors there are systematical effects due to finite volume and resolution of the simulation setup as well as the nonlinear nature of gravity. These errors are more difficult to  quantify and we will focus on providing estimates of how they can be minimised to sub-percent level.

We use {\tt Pkdgrav3} to run a suite of numerical simulations with varying box size and particle number. Volume effects on the power spectrum are investigated by comparing simulations with the same mass resolution and different box sizes. Effects due to particle numbers are studied with runs of constant box sizes. All simulations are based on 2LPT initial conditions, generated with {\tt MUSIC} at redshift 49. The power spectrum is measured with the triangular shaped cloud (TSC) mass assignment on a $8192^3$ grid.

\begin{figure}
\centering
\includegraphics[trim = 8mm 5mm 40mm 5mm, clip, scale=0.49]{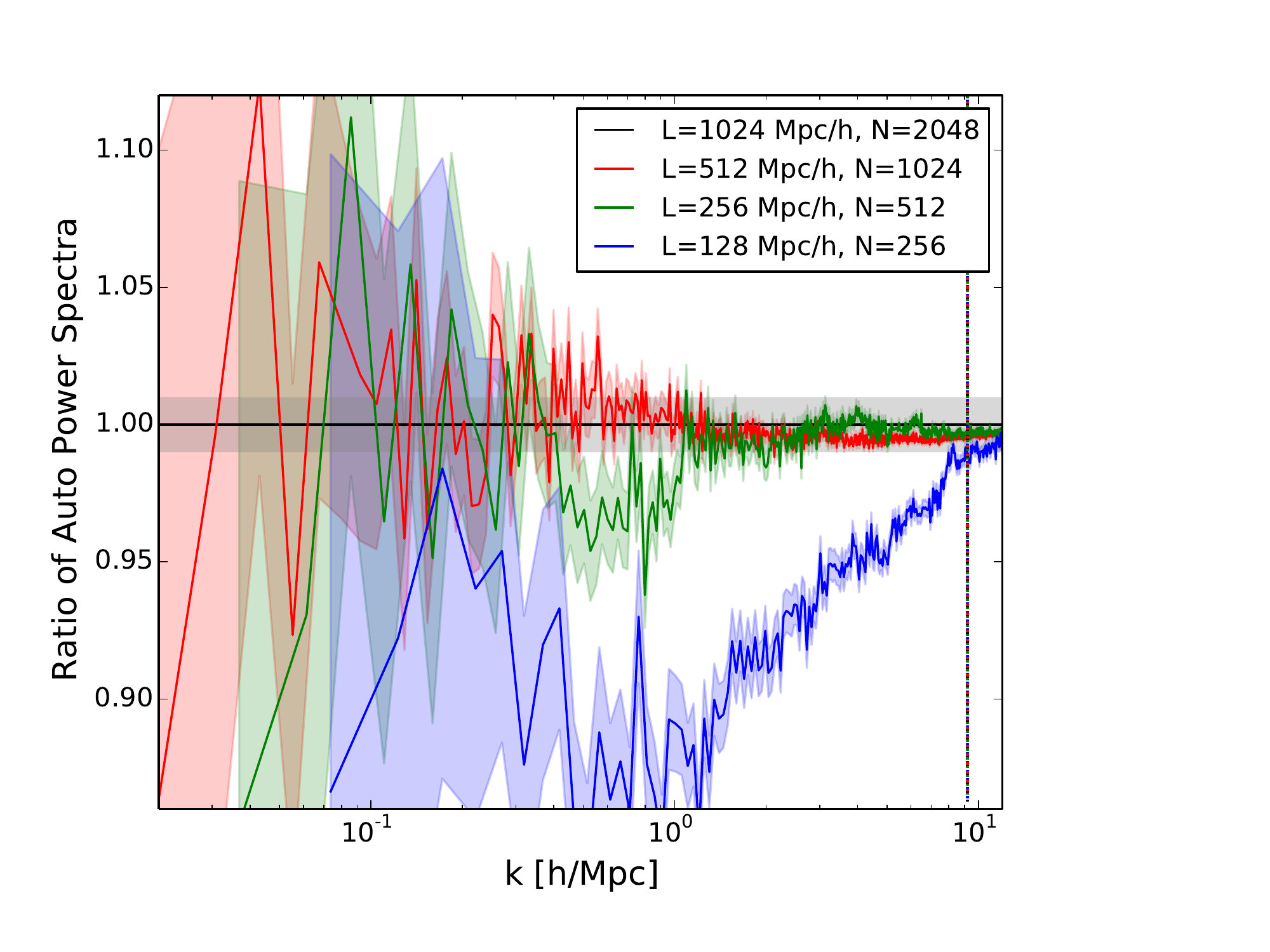}
\includegraphics[trim = 8mm 5mm 40mm 5mm, clip, scale=0.49]{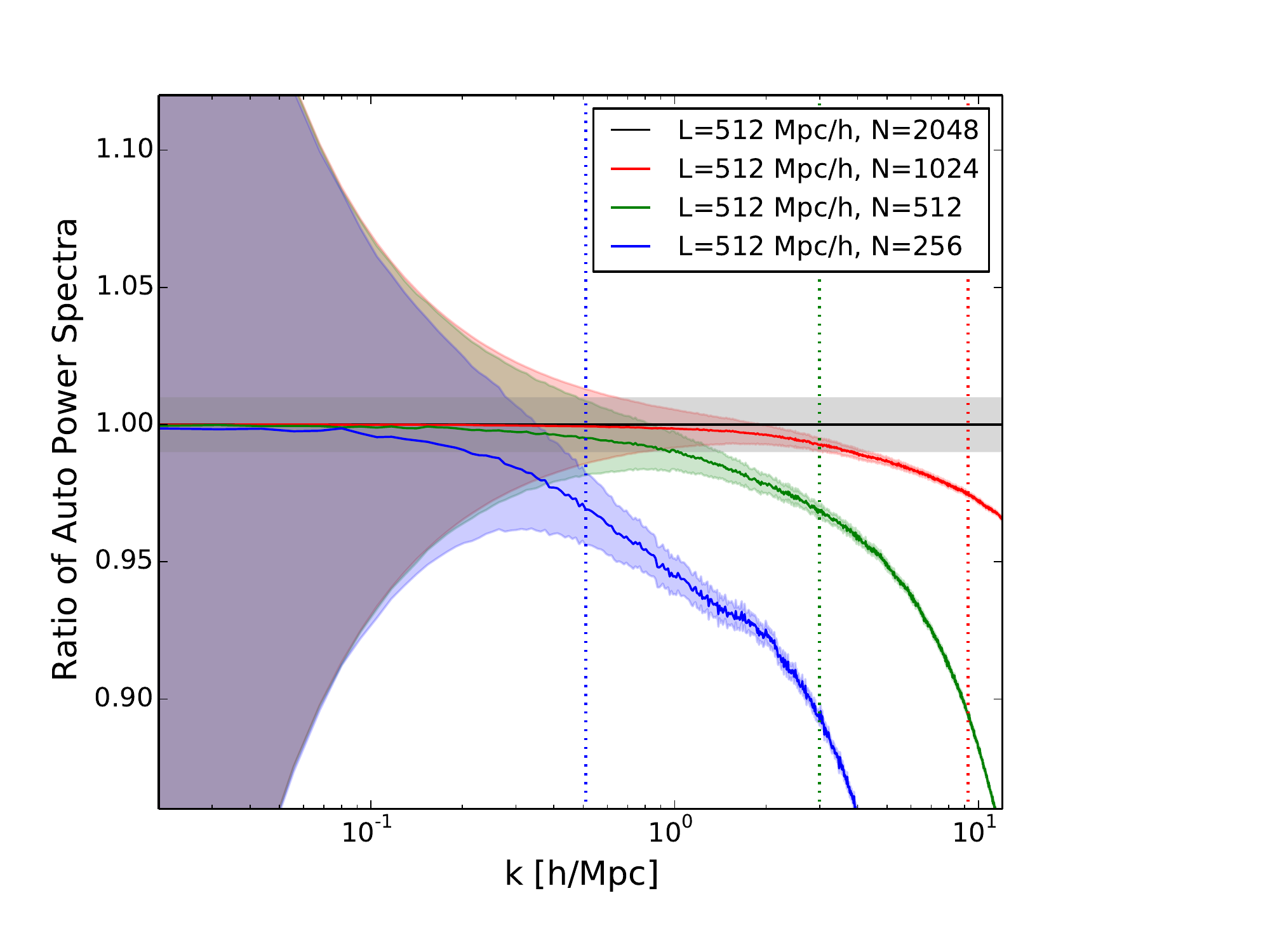}
\caption{\label{fig:PSbox}Investigating box-size and resolution effects on the power spectrum. {\it Left panel:} Same physical resolution and increasing box size:  L=128 $h^{-1}\rm Mpc$ (blue), L=256 $h^{-1}\rm Mpc$ (green), L=512 $h^{-1}\rm Mpc$ (red), L=1024 $h^{-1}\rm Mpc$ (black, reference line). {\it Right panel:} Same box size (L=512 $h^{-1}\rm Mpc$) and increasing number of particles (per dimension): N=256 (blue), N=512 (green), N=1024 (red), and N=2048 (black, reference line). The coloured areas quantify the statistical errors from sample variance, while the grey shaded band highlights the range of percent accuracy. Dotted vertical lines indicate where the shot-noise contribution exceeds one percent.}
\end{figure}

In the left panel of Fig.~\ref{fig:PSbox}, we illustrate ratios of power spectra from runs with the same mass resolution but different box sizes and particle numbers at redshift zero. The particle mass is kept constant at $M_p\sim10^{10}$ $h^{-1}$M$_{\odot}$, while the box length is increased together with the number of particles. A small box with $L=128$ $h^{-1}\rm Mpc$ (blue line) systematically underestimates the power by more than 10 percent. Doubling the box size to $L=256$ $h^{-1}\rm Mpc$ box (green line) leads to an overall accuracy of 5 percent (one percent at small scales, $k>1$ $h\,\rm Mpc^{-1}$). Boxes with length of $L=512$ $h^{-1}\rm Mpc$ (red line) and more ($L=1024$ $h^{-1}\rm Mpc$, black reference line) differ by about one percent or less over the entire range of wave numbers. We therefore conclude that the box size of simulations should not be smaller than 500 $h^{-1}\rm Mpc$ in order to eliminate all systematic {\it nonlinear} finite-volume effects at the required percent precision (see also \citep{Gnedin2011,Mohammed2014b} for similar conclusions). Reducing the sample variance below the percent level over the entire $k$-range illustrated in Fig.~\ref{fig:PSbox} would require a much larger box of $L\sim12$ $h^{-1}\rm Gpc$.

In the right panel of Fig.~\ref{fig:PSbox}, we plot power spectra from runs with the same box size ($L=512$ $h^{-1}\rm Mpc$) and different particle numbers, effectively increasing the mass resolution. Simulations with $N=256$ (blue line), $N=512$ (green line), and $N=1024$ (red line) underestimate the power on small scales with respect to the $N=2048$ reference run (black line). The convergence rate with respect to the percent accuracy requirement is directly proportional to the scale where the simulation shot-noise becomes relevant, i.e. the wave number  $k_{\rm sn}$ at which $P_{\rm sn}/P\equiv 0.01$. (illustrated by the dotted vertical lines in Fig.~\ref{fig:PSbox}). The maximum wave number $k_{\rm max}$ warranting percent accuracy is well described by $k_{\rm max}=k_{\rm sn}/3$. For the runs shown in the right panel of Fig.~\ref{fig:PSbox} with $N=256,\,512,\,1024$ (blue, green and red lines) this results in $k_{\rm max}= 0.2,\,1.0,\,4.0$ $h\,\rm Mpc^{-1}$. Previous investigations by Refs.~\citep{Joice2009,Heitmann2010} have proposed $k_{\rm max}$ to be half the Nyquist frequency ($k_{\rm Ny}=\pi N/L$) instead. For the same runs this would lead to $k_{\rm max}= 0.78,\,1.58,\,3.14$ $h\,\rm Mpc^{-1}$ which does not exactly reproduce our results\footnote{Assuming a power-law dependence of the power spectrum, $P(k)\propto k^{-\alpha}$, the scaling of $k_{\rm max}=k_{\rm sn}/3$ goes as $k_{\rm max}\propto(N/L)^{3/\alpha}$. For the asymptotic limit of $\alpha=3$ both approaches - the convergence scale to be tied to the shot noise or to the Nyquist frequency -  yield the same scaling with $k$. For $\alpha<3$, however, $k_{\rm sn}$ converges somewhat faster, which is in better agreement with our simulations.}.

The drop in power of low resolution runs, visible in the right panel of Fig.~\ref{fig:PSbox}, can be explained in terms of analytical considerations: experiments with the halo model show that clusters significantly contribute to the power spectrum, while the presence of small haloes below $10^{11}$ $h^{-1}\rm M_{\odot}$ have a negligible effect \citep{Schneider2012,vanDaalen2015}. Since the simulations with lower resolution (represented by the coloured dots) do not resolve haloes down to masses of $10^{11}$ $h^{-1}\rm M_{\odot}$, they underestimate the physical power at small scales. The $N=2048$ simulation on the other hand, has a particle mass of $M_{p}\sim10^{9}$ $h^{-1}\rm M_{\odot}$, resolving $10^{11}$ $h^{-1}\rm M_{\odot}$ haloes with $\sim 100$ particles. Moreover, the convergence rate in the plot suggests that the $N=2048$ run is one percent accurate until $k\sim10$ $h\,\rm Mpc^{-1}$ at redshift zero.

Based on the right-hand-side panel of Fig.~\ref{fig:PSbox}, we can determine a minimal mass resolution required to obtain percent convergence in the matter power spectrum. Since the runs illustrated by the blue, green, and red lines underestimate the power by more than a percent for values above $k=0.25,\,1,\,4$ $h\,\rm Mpc^{-1}$, we can safely expect the black line to depart from the true answer beyond $k=10$ $h\,\rm Mpc^{-1}$. A conservative estimate therefore yields a maximum simulation particle mass of $M_{\rm p}=10^{9}$ $h^{-1}$M$_{\odot}$ guaranteeing a percent converged power spectrum at all scales up to $k=10$ $h\,\rm Mpc^{-1}$. This requirement can be relaxed to $M_{\rm p}\sim8\times10^{10}$ $h^{-1}$M$_{\odot}$ for wave numbers up to $k=1$ $h\,\rm Mpc^{-1}$ (as shown by the green line). Numerical simulations for upcoming survey missions need large boxes of at least 4 $h^{-1}\rm Gpc$ to cover the entire survey volume \citep{Ivezic2008,Laureijs2011}. This means that at least $N=16000$ particles per dimension (i.e four trillion in total) are required to reach percent precision for the power spectrum up to $k\sim10$ $h\,\rm Mpc^{-1}$.

\subsection{Best guess for the power spectrum}
After investigating the convergence with respect to box size and mass resolution, we present a suite of four simulations with each $N=2048$ per dimension and decreasing box sizes of $L=4096,\,2048,\,1024$, and $512$ $h^{-1}\rm Mpc$. These simulations provide a combined measurement of the power spectrum over the entire range of scales from $k\sim0.05$ $h\,\rm Mpc^{-1}$ to $k\sim10$ $h\,\rm Mpc^{-1}$.

\begin{figure}
\centering
\includegraphics[trim = 4mm 10mm 70mm 12mm, clip, scale=1.0]{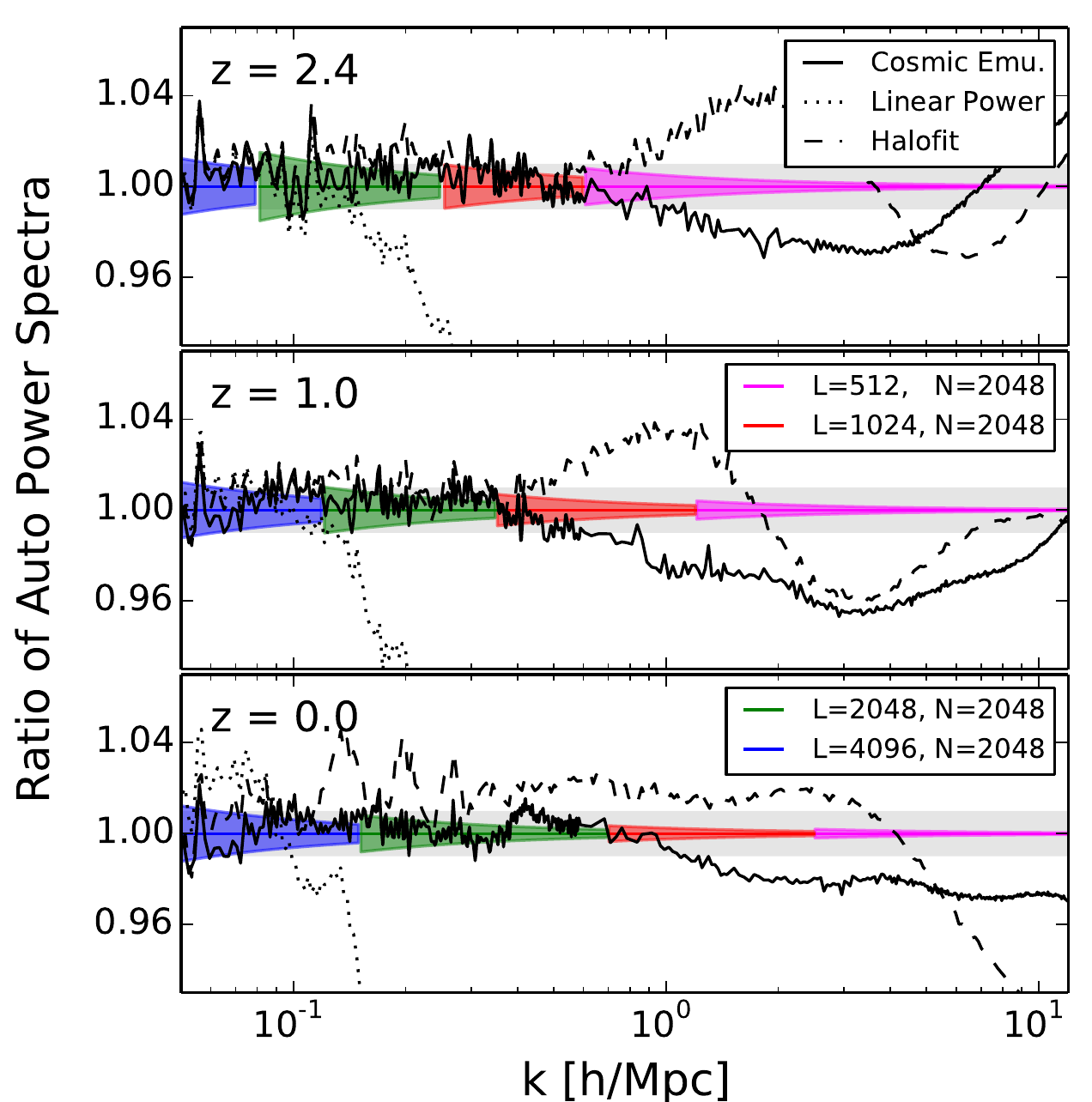}
\caption{\label{fig:PSemu}Power spectra of the cosmic Franken emulator \citep{Heitmann2014}, the revised halofit function \citep{Takahashi2012}, and the linear prediction compared to the outcome of simulations with $N=2048$ particles per dimension and varying box size $L$=512 $h^{-1}\rm Mpc$ (magenta), $L$=1024 $h^{-1}\rm Mpc$ (red), $L$=2048 $h^{-1}\rm Mpc$ (green), $L$=4096 $h^{-1}\rm Mpc$ (blue). Coloured areas quantify the statistical errors from sample variance and shot-noise, the grey shaded band highlights the range of percent accuracy.}
\end{figure}

In Fig.~\ref{fig:PSemu} we use the combined power spectrum from these four simulations as reference line, where the different colours indicate which simulation as been used for which k-range. The colour-shaded areas furthermore indicate the uncertainties due to sample variance and the grey-shaded area delimits the range of percent precision. In Fig.~\ref{fig:PSemu} the Franken emulator \citep[solid black line]{Heitmann2008}, the halofit model \citep[dashed black line]{Takahashi2012}, and the linear theory (dotted black line) are plotted against the reference line from our simulations.
The halofit model is a revised version of the \citet{Smith2003} fitting scheme, which is physically motivated by the halo model and claims to be 10 percent accurate for $k\leq1$ $h\,\rm Mpc^{-1}$ between $z=0$ and $z=10$. Compared to our simulations the agreement is better than 5 percent over all measured scales and redshifts. The cosmic Franken emulator is an interpolation tool based on a suite of simulations with varying cosmological parameters \citep{Heitmann2010,Heitmann2014}. The agreement between the emulator and our simulations is about three percent at $z=0$ and five percent at $z=1$. This is roughly within the stated accuracy of \citet{Heitmann2014}. However, our simulations consistently predict more power than the cosmic emulator at scales around $k\sim1$ $h\,\rm Mpc^{-1}$ and above, confirming results from \citet{Skillman2014} who observe a similar departure from the cosmic emulator. Part of the difference should come from the fact that the emulator was calibrated with {\tt Gadget} runs, while we use {\tt Pkdgrav}, two codes that differ by about three percent at $k>1$ $h\,\rm Mpc^{-1}$ as illustrated in Fig.~\ref{fig:PScomp}.


\section{Conclusions}\label{Conclusions}
The future of cosmology relies on data from large scale structure surveys. This data can only be fully exploited if we understand gravitational clustering and galaxy formation at high accuracy. The matter power spectrum, as the prime statistical measure, needs to be known within percent precision from linear scales up to $k\sim10$ $h\,\rm Mpc^{-1}$.

Although cosmological $N$-body techniques have been developed and constantly improved during the last two decades, obtaining the required accuracy remains a challenge. The entire pipeline from the generation of initial conditions to the analysis of the final data needs to be examined carefully and potential sources of error have to be quantified.

In this paper, we compare power spectra of simulations from the three gravity codes {\tt Ramses}, {\tt Pkdgrav3}, and {\tt Gadget3}. These codes are well established in the community and represent common $N$-body techniques for cosmological simulations: the particle-mesh technique, the tree method, and a hybrid combination of the two. In a second part, we explore potential error sources from {\it initial conditions}, {\it simulation volume}, and {\it resolution}, investigating effects on the matter power spectrum. These findings are then expressed in terms of a minimal volume and minimal mass resolution requirement to obtain the targeted percent accuracy.

The main results of the paper can be summarised as follows:
\begin{enumerate}

\item {\it Gravity calculation:} The gravity codes  {\tt Ramses}, {\tt Pkdgrav3}, and {\tt Gadget3} agree within one percent up to $k=1$ $h\,\rm Mpc^{-1}$ (over all studied redshifts), and within three percent up to $k=10$ $h\,\rm Mpc^{-1}$ (below redshift one). Increasing the accuracy of the global time integration is likely to further reduce errors at the largest scales (as suggested by our analysis of the phase spectrum and the bispectrum). Things are likely to be much more challenging at small scales, as there is no reference solution for the nonlinear power spectrum.

\item{\it Simulation volume:} A box size larger than $L\sim0.5$ $h^{-1}\rm Gpc$ is needed to avoid biases from nonlinear finite-volume effects at the percent level of the power spectrum. Reducing the Gaussian sample variance to a sub-percent level for the $k$-range above $k=0.1$ $h\,\rm Mpc^{-1}$ would require an even larger box size of $L\sim2.5$ $h^{-1}\rm Gpc$.

\item{\it Mass resolution:} A conservative estimate of the maximum particle mass in simulations yields $M_{\rm p}=10^{9}$ $h^{-1}\rm M_{\odot}$ for percent accurate power spectra up to $k=10$ $h\,\rm Mpc^{-1}$. This requirement can be relaxed to $M_{\rm p}\sim8\times10^{10}$ $h^{-1}\rm M_{\odot}$, if only wave numbers up to $k=1$ $h\,\rm Mpc^{-1}$ are considered. Upcoming surveys, such as {\tt DES}, {\tt LSST}, and {\tt Euclid}, require large simulation volumes of $L\sim4$ $h^{-1}\rm Gpc$ or more. As a consequence, numerical simulations need to have at least $N\sim16000$ particles per dimension (i.e four trillion in total) to reproduce the power spectrum at targeted accuracy.

\item{\it Initial conditions:} Initial conditions based on the Zel'dovich approximation (ZA) require very high starting redshifts of $z_i=200$ or above. Such high redshifts are prone to numerical errors, since the size of perturbations are of the order of the numerical accuracy. Initial conditions based on second order Lagrangian perturbation theory (2LPT) are significantly more accurate. They allow late starting redshifts, reducing the run-time of simulations and minimising potential numerical errors in the high redshift regime.
\end{enumerate}

Summarising these results, it is possible to run cosmological simulations with sub-percent errors from volume and mass resolution effects, however, at the price of very high particle numbers. In terms of the gravity calculation, the agreement between codes is good, but not quite at the percent level for the very nonlinear regime.

In the future, it will be crucial to include baryonic effects driven by AGN feedback, as they have been shown to significantly affect the matter power spectrum at nonlinear scales \citep{vanDaalen2011,Angulo2013}. Quantifying and parametrising the AGN feedback will be one of the main challenges of computational cosmology, and a basic requirement to take full advantage of the upcoming large scale structure observations.

\section*{Data Release}
All relevant data of the code comparison project, i.e. the IC file, run parameters, and power spectra measurements, can be found at {\tt www.ics.uzh.ch/{\raise.17ex\hbox{$\scriptstyle\sim$}}aurel/}. We hope that this information will be useful for future comparison and accuracy tests including other $N$-body codes. 

\section*{Acknowledgements}
This work was initiated within the framework of the Cosmological Simulation Working Group of the Euclid Consortium. AS is supported by the Synergia project ``Euclid'' from the Swiss National Science Foundation. VS acknowledges support from the Deutsche Forschungsgemeinschaft (DFG) through Transregio 33, ``The Dark Universe''. All simulations were run on the {\it Piz Daint} cluster at the Swiss National Supercomputer Centre (CSCS) under the project allocation s511.


\bibliographystyle{plain}


\appendix
\label{Appendix}
\section{Variation of code parameters}\label{ApdxA}
In the main text, we use standard code parameters from the literature to compare the different gravity codes. This is justified because the exact solution of the power spectrum is not known at nonlinear scales, and a posterior adjustment of code parameters would lead to a false impression of convergence. It is nevertheless important to quantify how the choice of code parameters affects the final results.

In general, code accuracy parameters can either be attributed to the force calculation or the time-stepping. Typical parameters regulating the force accuracy are softening-length and opening-angle for tree-codes (such as {\tt Pkdgrav3}) as well as grid refinement strategy and accuracy of the Poisson solver for adaptive PM codes (such as {\tt Ramses}). Hybrid codes (such as {\tt Gadget3}) usually have an additional parameter regulating the transition scale between the PM and tree regime (PM-grid). The accuracy of time integration, on the other hand, is usually controlled by the adaptive time-stepping criterion which is implemented in a similar way in all three codes.

Past work has shown that for tree codes softening and tree-opening criteria show percent convergence at $k\leq10$ $h\,\rm Mpc^{-1}$for reasonable parameter choices \citep{Reed2013,Smith2014}. The same seems true for the force accuracy parameters of adaptive PM codes which have shown to yield the same precision than generic tree-codes \citep{Teyssier2002}. More significant deviations are reported for the transition parameter between the PM and tree regimes in hybrid codes \citep{Smith2014} and for the time-stepping criterion affecting all three codes \citep{Reed2013}. In the following, we investigate the effects of both time-stepping and PM-grid transition on the resulting power spectrum.

\begin{figure}
\centering
\includegraphics[trim = 7mm 10mm 70mm 12mm, clip, scale=0.525]{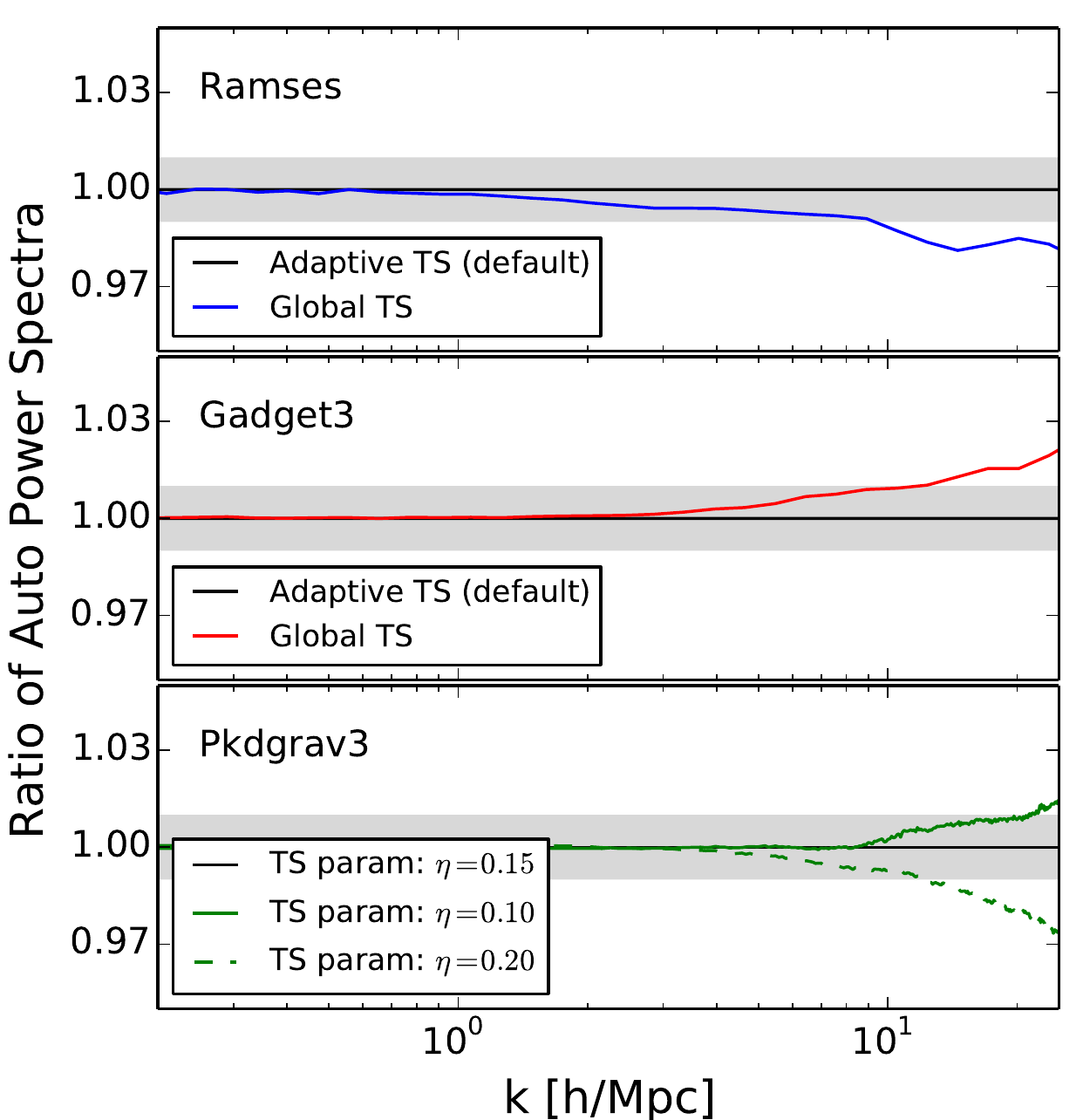}
\includegraphics[trim = 7mm 10mm 40mm 12mm, clip, scale=0.525]{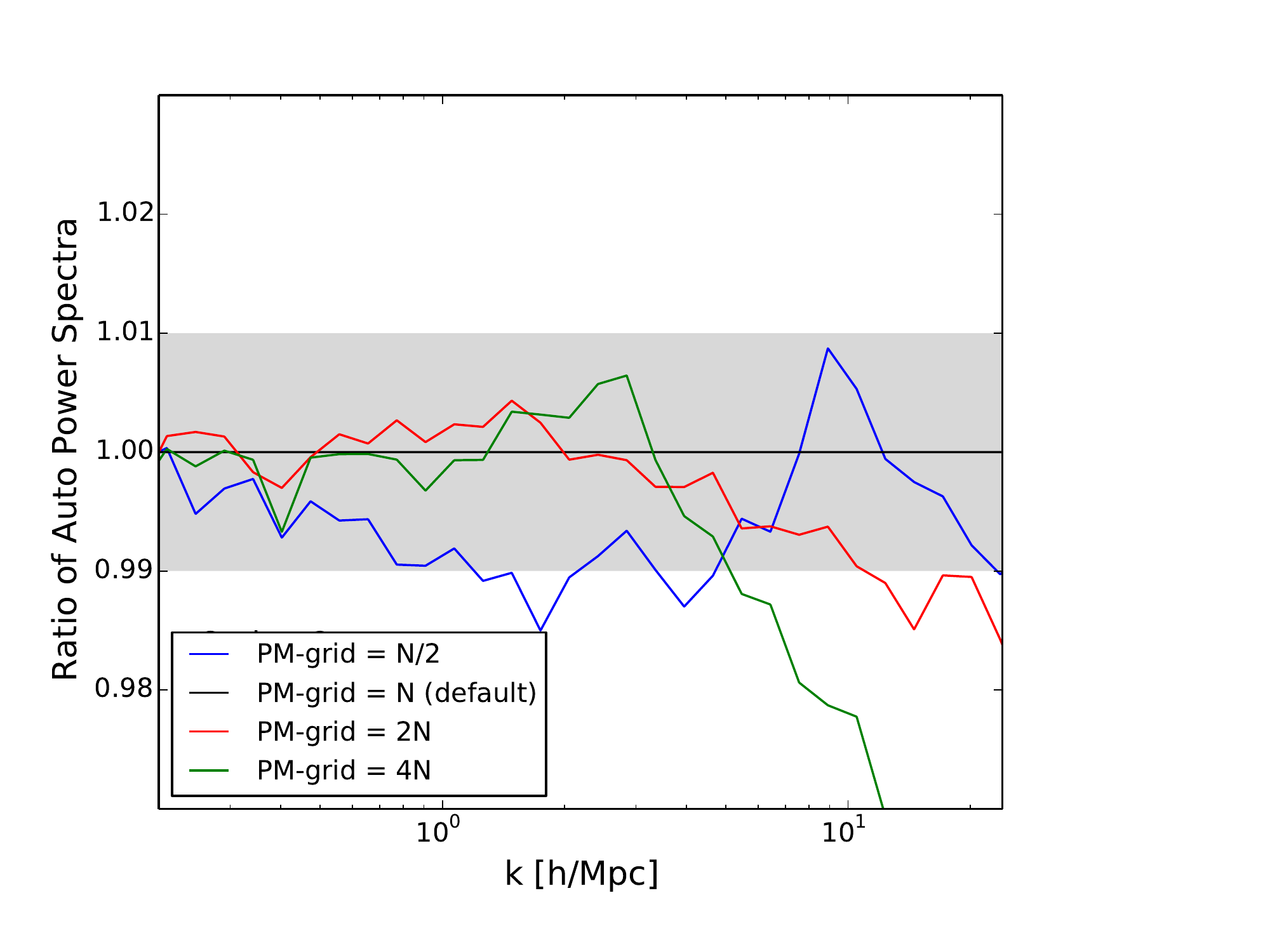}
\caption{\label{fig:PSaltTS}{\it Left:} Different time-stepping strategies and their effects on the auto power spectrum. For {\tt Ramses} (top) and {\tt Gadget3} (centre) we show adaptive (default) and global time-stepping, for {\tt Pkdgrav3} (bottom) we vary the  time-stepping parameter $\eta$ around the default value $\eta=0.15$. {\it Right:} Varying grid size of the PM-mesh in {\tt Gadget3} and how this affects the auto power spectrum.}
\end{figure}

In order to test the effect of time-stepping, we run simulations with alternative time-stepping criteria for all three codes of the comparison project. For {\tt Ramses} and {\tt Gadget3} we use the global time-stepping mode as an alternative, where {\it all} particles trajectories are integrated with the smallest time-stepping of the adaptive (default) mode independently of their gravitational acceleration. For {\tt Pkdgrav} we keep the adaptive nature of time-stepping and vary the time-stepping parameter $\eta$ around the default choice $\eta=0.15$.

The impact of the time-stepping on the power spectrum is illustrated in the left panel Fig.~\ref{fig:PSaltTS}. Switching to global time-stepping affects the result at the percent level around $k\sim10$ $h\,\rm Mpc^{-1}$ for both {\tt Ramses} and {\tt Gadget3}, however, with an inverse general trend (reducing power for {\tt Ramses} and increasing it for {\tt Gadget3}). Varying the $\eta$ parameter in {\tt Pkdgrav3} around $\eta=0.1-0.2$ also leads to a percent effect on the power spectrum at $k\sim10$ $h\,\rm Mpc^{-1}$ (bottom panel), with the general trend of increasing power for smaller time-steps. Based on these tests we conclude that the results from the code comparison (i.e., Fig.~\ref{fig:PScomp}) are not sensitive to the time-step criterion as long as reasonable parameter choices are considered.

The effect of the PM-grid transition in {\tt Gadget} is investigated by running simulations with different size of the PM grid around the default choice (where the number of grid points equals the particle number, i.e., PM-grid $= N$). The resulting power spectra are illustrated in right panel of Fig.~\ref{fig:PSaltTS}, showing differences at the percent level over various scales. The size of the scatter seems significant for precision cosmology and requires further investigation. However, the variation is not large enough to explain the offset between {\tt Gadget} and {\tt Ramses}/{\tt Pkdgrav} in the $z=0$ panel of Fig.~\ref{fig:PScomp}.

We have shown in this appendix that changing the time-stepping criterion of our codes has a sub-percent effect on the auto power spectrum below $k\sim10$ $h\,\rm Mpc^{-1}$. The error induced by the PM-tree-transition in {\tt Gadget} is slightly larger but still roughly below one percent. As argued above, other parameters, such as softening and tree opening for tree-codes as well as the accuracy of the Poisson solver for mesh-codes are expected to yield even smaller errors. We therefore conclude that simple tuning of parameters is not enough to bring the different codes into sub-percent agreement. Deeper investigations of the discretisation and integration techniques might be required to achieve this goal.

\section{Beyond the power spectrum: propagator and bispectrum}\label{ApdxB}
In this appendix we consider a different set of statistics from the main text to have a deeper understanding of the differences between the $N$-body codes and also illustrate the robustness of the results obtained so far. The first is the propagator $G(k)$ which results from the cross-correlation of the initial conditions $\delta_0$ (common to all codes) with the density fluctuations $\delta$ at a given redshift,
\begin{equation}
  G(k,z) \equiv {\langle \delta(\mathbf{k},z)\, \delta_0(\mathbf{k}')\rangle  \over \langle\delta_0(\mathbf{k})\, \delta_0(\mathbf{k}')\rangle}.
\end{equation}
This two-point statistic is sensitive to the displacement of particles away from their initial conditions on scales significantly larger than $2\pi/k$, unlike the equal-time power spectrum considered in the main text. At leading order in perturbation theory valid at large scales $G(k)$ agrees with the growth factor, whereas at small scales  the propagator drops to zero on scales smaller than the inverse of the rms displacement field at the given redshift \citep{CroSco2006}.

Since we have only one realisation for each $N$-body code, it is difficult to conclude anything from comparing the propagators to their expected large-scale limit, e.g. computed using RegPT \citep{BerCroSco1206}. For example, looking at the first five bins in Fourier space, all three codes agree with the predictions to better than $0.5\%$ at $z=2$ but there is no clear winner in terms of best agreement as the measurements fluctuate about the theoretical result as $k$ changes. For this reason it is more robust to look at the {\em ratio} between different codes and explore to what extent the differences between them can be understood. 

\begin{figure}
\centering
\includegraphics[width=0.45 \linewidth]{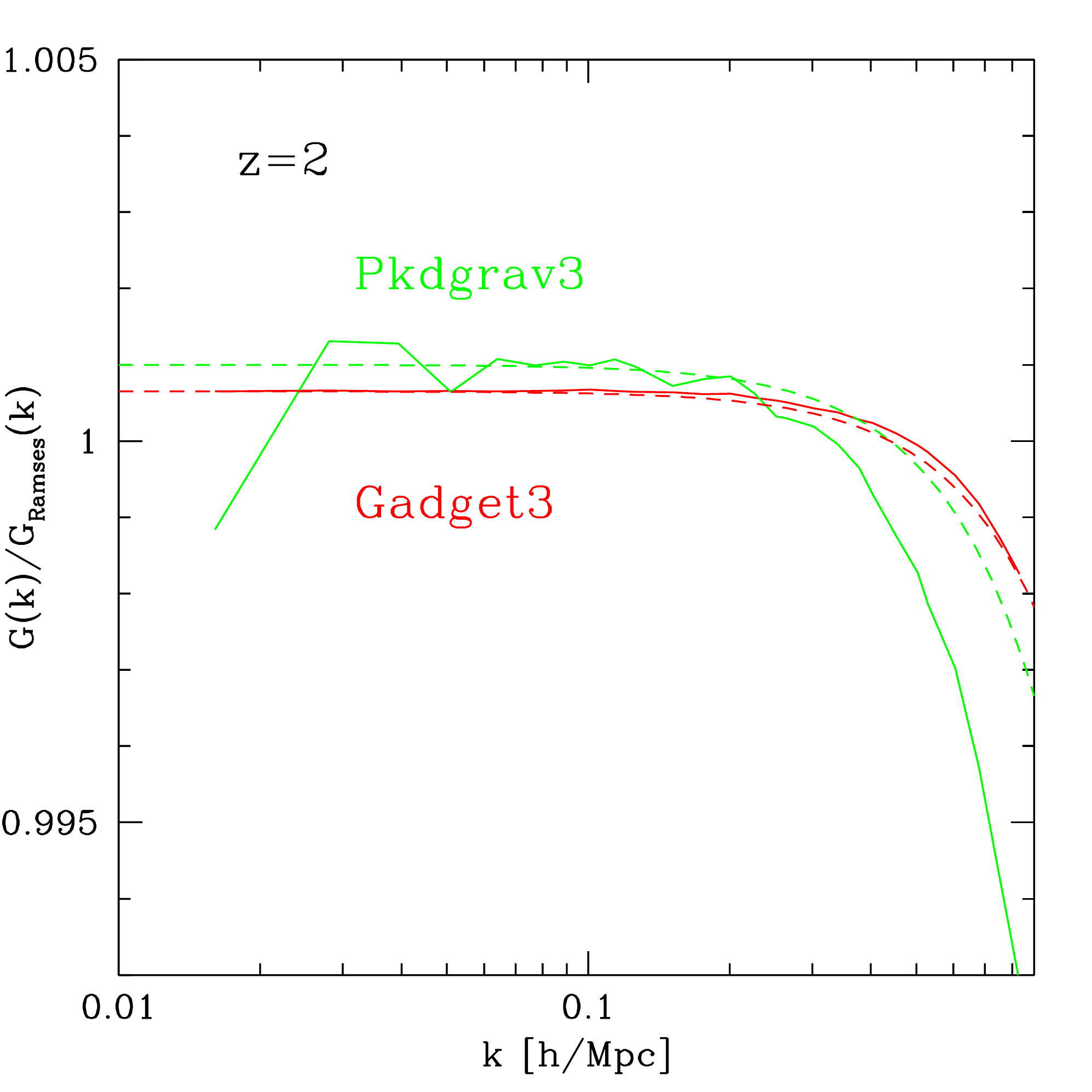}
\includegraphics[width=0.45 \linewidth]{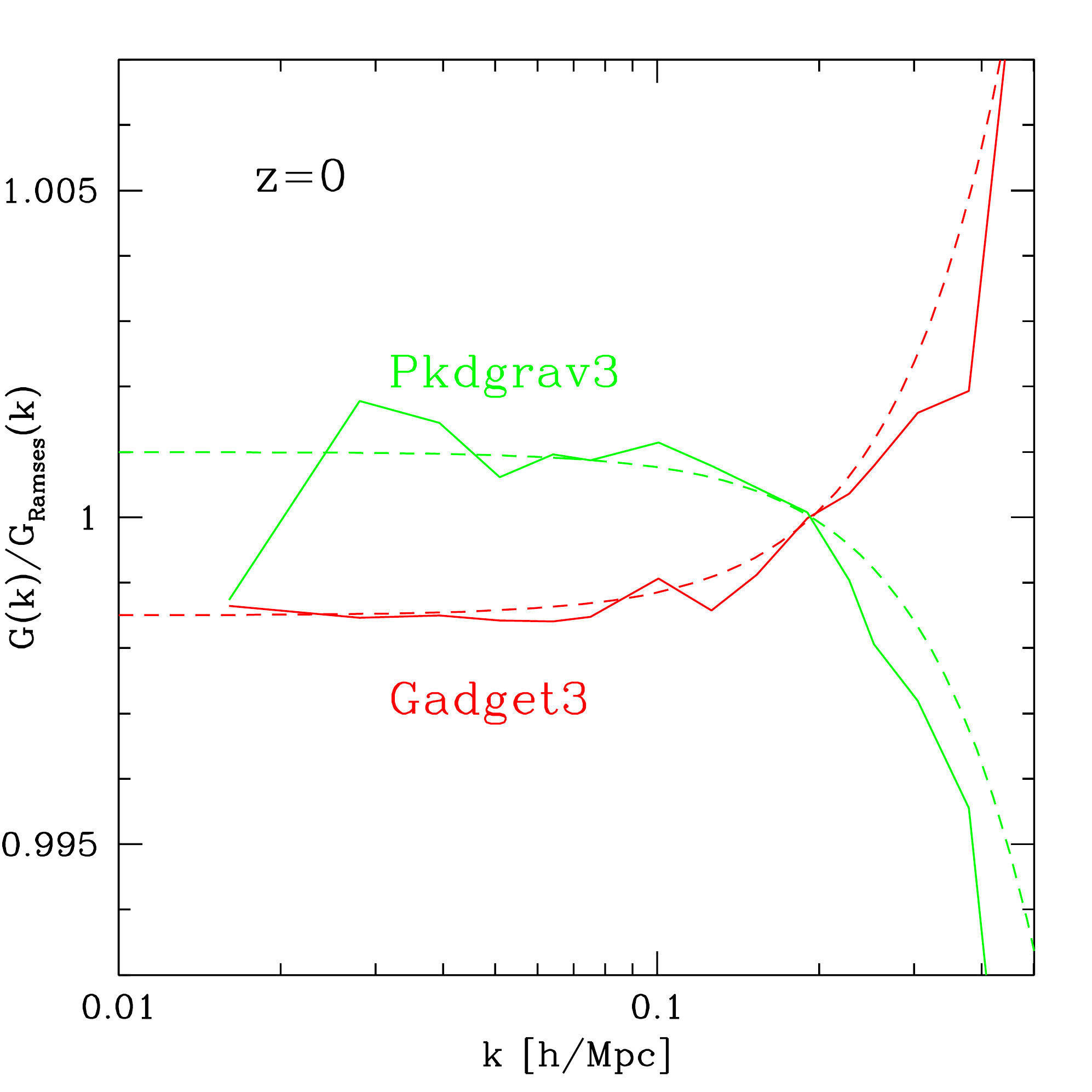}
\caption{\label{fig:Gk}{\it Left:} The solid lines show the ratio of {\tt Gadget3} and {\tt Pkdgrav3} propagators to the {\tt Ramses} propagator at $z=2$. The dashed lines denote the expectation of these ratios based on RegPT. {\it Right:} Same as left panel, but at  $z=0$.}
\end{figure}

In Fig.~\ref{fig:Gk} we show the ratio of the measured propagators to that of the {\tt Ramses} code, at $z=2$ (left panel) and $z=0$ (right panel), shown from the fundamental mode  of the simulation box up to scales where the propagator drops (exponentially) to zero. There are three points worth making here. First, the low-$k$ deviations in the propagator ratios are largely consistent (half the value) with those seen in the power spectrum in Fig.~\ref{fig:PScomp}. The second point is that a low-$k$ enhancement (suppression) goes together with a high-$k$ suppression (enhancement) of the propagators. This makes sense as a low-$k$ enhancement corresponds to an overall larger displacement field, which also decreases the propagator at small scales as the cross-correlation between initial and final conditions is suppressed by what is, effectively, slightly more time evolution. Finally, the redshift dependence shows that while the relative behaviour of {\tt Pkdgrav3} and  {\tt Ramses} is the same at $z=0$ and $z=2$, the {\tt Gadget3} deviations compared to {\tt Ramses} have opposite signs at the outputs considered here.

The figure also shows the expected propagator ratios from RegPT (dashed lines). At $z=2$ the expectation works very well for the ratio of {\tt Gadget3} to {\tt Ramses} propagators, but not so well for the {\tt Pkdgrav3} to {\tt Ramses}, which also shows significant more noise, particularly at the fundamental mode of the box. Clearly, the early (up to $z=2$) time evolution of {\tt Gadget3} and {\tt Ramses} are consistent with each other except for some small (less than $0.1\%$) relative displacement, while the evolution of {\tt Pkdgrav3}  is not as consistent with {\tt Ramses} (or {\tt Gadget3}) in terms of an overall slight displacement mismatch (as shown by comparison with RegPT and the relative fluctuations. This is perhaps not surprising, as the large-scale forces are computed by the PM method in both {\tt Ramses} and {\tt Gadget3}, while a tree is used in the {\tt Pkdgrav3} case. As the evolution proceeds to $z=0$, however, the relative evolution of {\tt Gadget3} to {\tt Ramses} drifts and changes sign (at $z=0$ {\tt Gadget3} is slightly less evolved than {\tt Ramses}) with this relative behavior still fairly well predicted by RegPT. For {\tt Pkdgrav3} the low-$k$ noise remains, but the relative evolution to {\tt Ramses} appears much more consistent than it was at $z=2$ to what is expected from RegPT. 

We now consider the bispectrum. Given the discussion above, we expect that the difference in the bispectra obtained from the different simulations to differ mostly by an overall constant growth factor, and since  to leading order in perturbation theory the bispectrum scales as the power spectrum squared, the difference in amplitude between bispectra should be about twice that observed in the power spectrum in Fig.~\ref{fig:PScomp} (or four times that in the propagator in Fig.~\ref{fig:Gk}). For this reason, it is convenient to show results for the {\em reduced bispectrum} $Q_{123}$ defined as,
\begin{equation}
  Q_{123} \equiv {B_{123} \over {P_1\, P_2 + P_2\, P_3 + P_3\, P_1 }},
\end{equation}
where $P_i\equiv P(k_i)$ and $ \langle\delta(\mathbf{k}_1)\, \delta(\mathbf{k}_2)\, \delta(\mathbf{k}_3)\rangle = \delta_D(\mathbf{k}_1+\mathbf{k}_2+\mathbf{k}_3) \ B_{123}$ with $B_{123}$ the bispectrum. The reduced bispectrum is, to leading order in perturbation theory,  independent of the overall value of the growth factor.

\begin{figure}[t!]
\centering
\includegraphics[width=0.45 \linewidth]{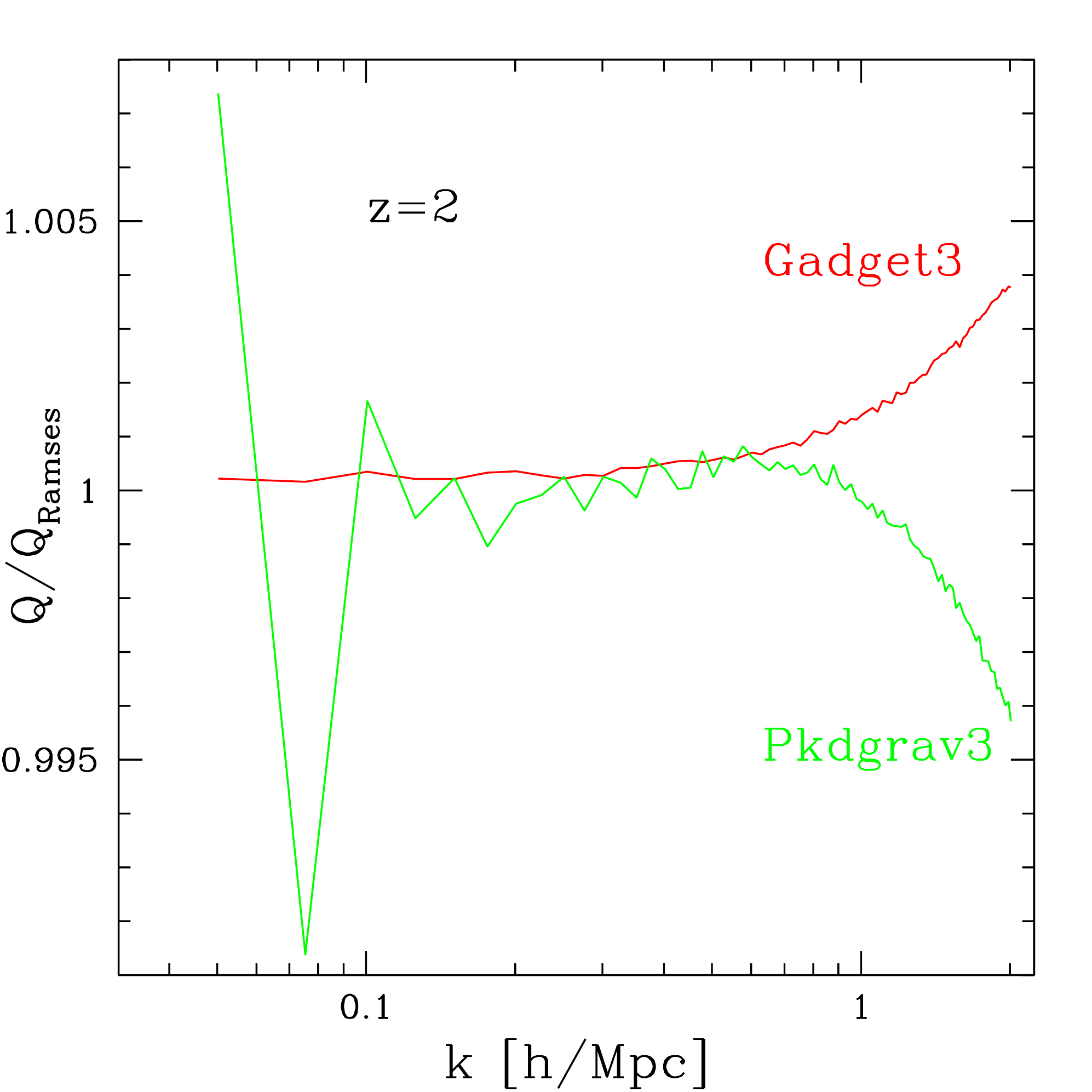}
\includegraphics[width=0.45 \linewidth]{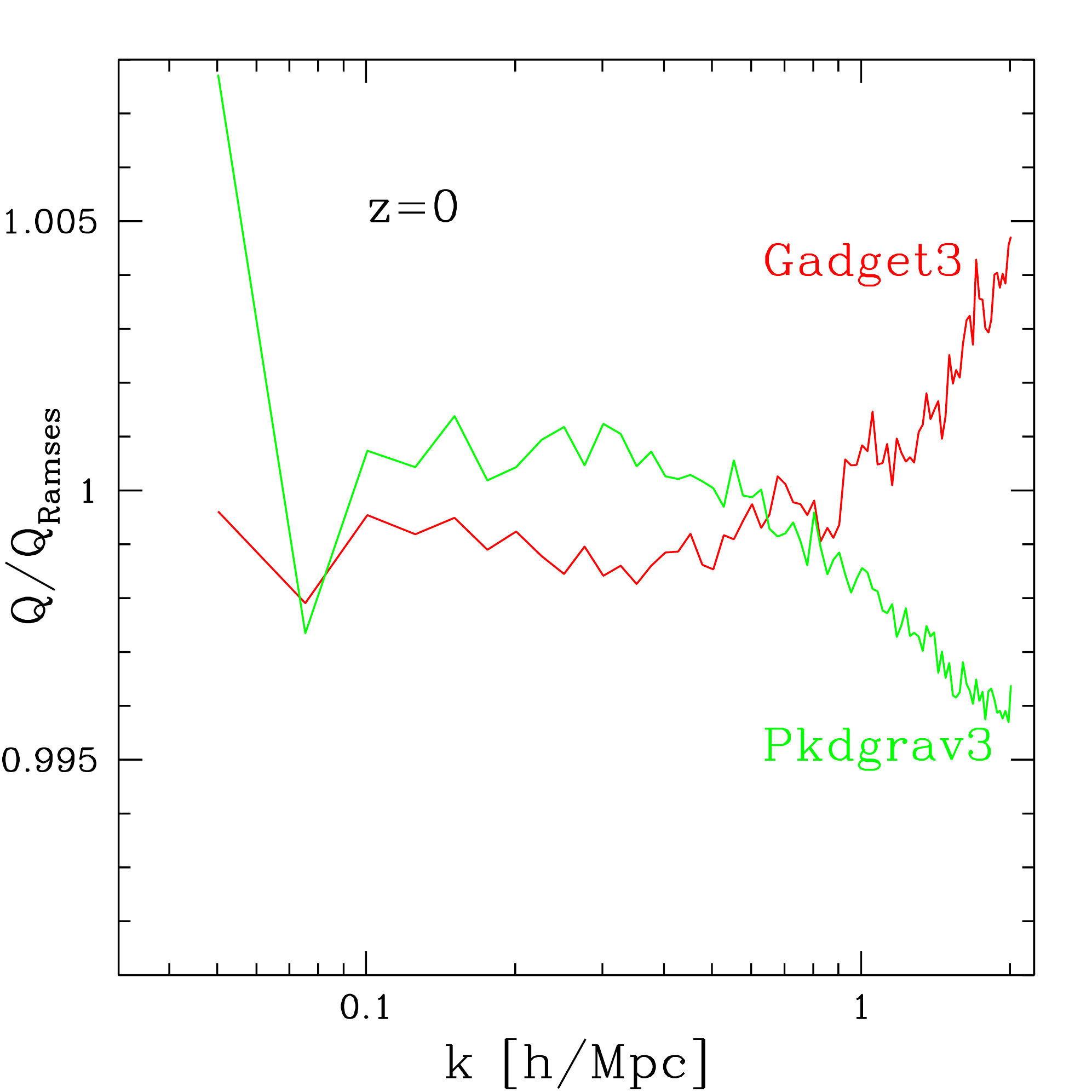}
\caption{\label{fig:Qk}{\it Left:} The ratio of {\tt Gadget3} and {\tt Pkdgrav3} reduced bispectra to the {\tt Ramses}  reduced bispectrum at $z=2$. The reduced bispectrum in each case corresponds to an average over all triangles with maximum wavenumber equal to $k$. {\it Right:} Same as left panel, but at  $z=0$.}
\end{figure}

Figure~\ref{fig:Qk} shows the results for the reduced bispectra at $z=0,2$. We have measured the bispectrum for all triangle shapes from scales of twice the fundamental mode of the box $k_f \simeq 0.0126$ $h\,\rm Mpc^{-1}$ up to $160\, k_f \simeq 2.01$ $h\,\rm Mpc^{-1}$, using the method discussed in \citep{Sefusatti2015}. For simplicity we show the results for the reduced bispectra after averaging all triangles whose maximum side is $k$ (the horizontal axis in Fig.~\ref{fig:Qk}). We see from this figure that the reduced bispectra are overall in remarkable agreement at the sub-percent level, while the agreement between {\em bispectra} (not shown) is at the one-percent level at large-scales (as expected from the power spectrum results and the scaling discussed above). In particular, we see that at $z=2$ the agreement between {\tt Gadget3} and {\tt Ramses} is essentially perfect until $k\sim 0.6$ $h\,\rm Mpc^{-1}$ and then it shoots up by only 0.4\% by $k\sim2$ $h\,\rm Mpc^{-1}$. On the other hand, for {\tt Pkdgrav3} the differences are more noticeable for the low-$k$ modes (as noted for the propagator and power spectrum before). At $z=0$, for most triangles at low-$k$ the differences remain at the $0.2\%$ level at most, while at high-$k$ the remain below $0.5\%$. Overall these results are consistent with the picture discussed above, that the evolution of {\tt Gadget3} and {\tt Ramses} are fairly consistent with each other up to an overall small mistmatch in growth factors, while the early time evolution of {\tt Pkdgrav3} differs by a bit more than just an overall scale factor at large scales ($k < 0.1$ $h\,\rm Mpc^{-1}$). In fact, for the sake of clarity Fig.~\ref{fig:Qk} starts from $k=4k_f$, but  the ratio of {\tt Pkdgrav3} to {\tt Ramses} reduced bispectra for equilateral triangle of sides $k=2k_f$ is as large as $\simeq 1.05$ at $z=0,2$. Unfortunately there is no reliable way of telling which results, {\tt Pkdgrav3} on the one hand, or {\tt Ramses} and {\tt Gadget3} (which are consistent among themselves), are the correct ones as the cosmic variance is significant at these scales for a single realization of a relatively small simulation box. At small scales all the codes differ in their reduced bispectra at below the one-percent level and the situation is even less clear cut. It seems, however, entirely possible that making the large-scale factors agree better than we have here can move towards making those discrepancies even smaller, as the reduced bispectrum is affected by overall growth in the nonlinear regime. To make this more quantitative, since the {\em rms} large-scale displacement at $z=0$ is of order 10 $h\,\rm Mpc^{-1}$, the error seen on it of order 0.1\% (see Fig.~\ref{fig:Gk}) corresponds to 0.01 $h\,\rm Mpc^{-1}$ errors on displacements, which of course can give larger than percent corrections to the power spectrum for $k>1$ $h\,\rm Mpc^{-1}$. In addition, the large-scale enhancement of displacements (or growth factors) which correspond to slightly more evolved configurations do have an enhancement of the small-scale power spectrum, as expected.

\end{document}